\documentclass[12pt]{article}

\usepackage{graphicx,epsfig,rotate,amsmath}    
\usepackage{latexsym,color,amssymb,citesort}
\newcommand{\be}{\begin{equation}}
\newcommand{\ee}{\end{equation}}
\newcommand{\eq}{\begin{eqnarray}}
\newcommand{\en}{\end{eqnarray}}
\newcommand{\bea}{\begin{eqnarray}}
\newcommand{\eea}{\end{eqnarray}}

\newcommand{\ed}{\end{document}}
\newcommand{\bc}{\begin{center}}
\newcommand{\ec}{\end{center}}

\begin{document}

\thispagestyle{empty}

\begin{flushright}
{\tiny Preprint HISKP--TH--10/21, FZJ-IKP(TH)--2010--18}
\end{flushright}

\begin{center}

\vspace{1.75cm}
{\Large{\bf Scalar mesons in a finite volume}}

\vspace{0.5cm}

\today

\vspace{0.5cm}

V. Bernard$^a$,
M. Lage$^b$,
U.-G. Mei{\ss}ner$^{b,c}$ and
A.~Rusetsky$^b$

\vspace{2em}

\begin{tabular}{c}
$^a\,${\it Groupe de Physique Th\'eorique,
Universit\'e de Paris-Sud-XI/CNRS,}\\{\it  F-91406 Orsay, France}\\[2mm]
$^b\,${\it Helmholtz--Institut f\"ur Strahlen-- und Kernphysik}\\
{\it and Bethe Center for Theoretical Physics, Universit\"at Bonn}\\
{\it  D--53115 Bonn, Germany}\\[2mm]
$^c${\it Forschungszentrum J\"ulich, J\"ulich Center for Hadron Physics,}\\
{\it Institut f\"ur Kernphysik  (IKP-3)
 and Institute for Advanced Simulation (IAS-4),}\\ {\it D-52425 J\"ulich, Germany}
\end{tabular}

\end{center}

\vspace{1cm}

{\abstract
{Using effective field theory methods, we discuss the extraction of 
the mass and width of the scalar mesons $f_0(980)$ and $a_0(980)$ from the
finite-volume spectrum in lattice QCD. In particular, it is argued that
the nature of these states can be studied by invoking twisted boundary
conditions, as well as investigating the quark mass dependence of the spectrum.  
}
}

\vskip1cm

{\footnotesize{\begin{tabular}{ll}
{\bf{Pacs:}}$\!\!\!\!$& 11.10.St, 11.15.Ha, 11.80.Gw, 13.75.Lb\\
{\bf{Keywords:}}$\!\!\!\!$& Resonances in lattice QCD,
field theory in a finite volume,\\$\!\!\!\!$ & scalar mesons, $\pi\pi$ scattering
\end{tabular}}
}
\clearpage


\section{Introduction}

The scalar sector of the low-energy QCD is controversial. 
In particular, in the experimental spectrum
there are too many candidates for the scalar $q\bar q$ nonet.
In the phenomenological approaches, alternative solutions to this 
problem have been suggested.
One of the possible solutions is to treat some of these mesons
as tetraquark states 
(see, e.g.,~\cite{Jaffe:1976ig,Black:1998wt,Achasov:2002ir,Pelaez:2004xp}).
Another suggestion is that $a_0(980)$ and
$f_0(980)$ are to be considered as $K\bar K$
molecules~\cite{Weinstein:1982gc,Oller:1997ng,oop}. Further, 
in Refs.~\cite{Oller:1998zr}
these states were described as a combination of a bare pole and 
the rescattering contribution. The investigations carried out within the
framework of QCD sum rules are, in particular,
 indicative of the non-$q\bar q$ nature of $a_0(980)$~\cite{sumrules}.
In the J\"ulich meson-exchange model, $f_0(980)$ appears to be a
 bound $K\bar K$ state, whereas $a_0(980)$ is a dynamically 
generated threshold effect \cite{Janssen:1994wn}.
In the view of the above controversial identifications  we wish to stress 
that all these, in general, are model-dependent and can not be 
unambiguously interpreted in quantum field theory. However, in case when
the states are very close to some 2-particle threshold (as it is
indeed the case with $a_0(980)$ and $f_0(980)$), it is possible
to make a model-independent statement, whether these resonances are 
molecular states or not. The ``compositeness criterium,'' which is applied
here, was first introduced by Weinberg~\cite{Weinberg:1963zz}. This approach
is related to the ``pole counting'' method, considered in 
Refs.~\cite{Morgan:1992ge,Tornqvist:1994ji}.
The above methods were used, e.g., in
Refs.~\cite{Morgan:1993td,Baru:2003qq,Baru:2004xg,Hanhart:2006nr,Hanhart:2007cm} to study the
nature of $a_0(980)$ and $f_0(980)$ resonances. In particular, in 
Ref.~\cite{Baru:2004xg}, the position of the $S$-matrix poles 
in the vicinity of the $K\bar K$ threshold in the scalar sector of QCD
is expressed through the so-called Flatt\'e parameters, which
describe a resonance located in the vicinity of a 2-particle threshold
and which are in principle measurable in the scattering experiments.
Note, however that the compositeness criterium (or the pole counting method)
is designed to distinguish a near threshold molecular state from a tightly
bound system of quarks. The question about the precise nature of this quark
compound ($q\bar q$ state, or tetraquark, or a glueball, or something else), can not
be resolved by this criterium.

It is often stated that the study of the scalar spectrum in lattice QCD 
can eventually lead to the understanding of the nature of these
states. Indeed, recent years have seen considerable activity, concerning
the calculation of the spectrum of the scalar mesons on the
lattice (see, e.g.,~\cite{Alford:2000mm,Kunihiro:2003yj,Suganuma:2005ds,Mathur:2006bs,Hart:2006ps,Wada:2007cp,Prelovsek:2010gm}).
However, alone the calculations of the excited spectrum 
do not answer the question. Additional criteria are usually applied.
It should be noted that, as compared to the phenomenological approaches,
 lattice QCD
has in general more tools at its disposal which can be
exploited in order to separate the exotic states from the conventional
$q\bar q$ spectrum.
 We mention, as one example, the method of hybrid boundary conditions 
(HBC)~\cite{Suganuma:2005ds}, which is used to distinguish the 
scattering states from the tightly bound quark-antiquark systems. Another example is given by the
 calculations of the spectrum of the $q^2\bar q^2$ mesons, which were
done in the quenched
approximation and in the absence of $q\bar q$ annihilation 
diagrams~\cite{Alford:2000mm}. In the latter paper it has been argued that,
due to the above approximations, and due to the fact that the quark masses
used in the simulations were rather high, mixing of the $q\bar q$ channel
to the tetraquark states is suppressed, so that the observed spectrum can be
readily attributed to the latter.
However,
even if these arguments might look intuitively plausible, they 
explicitly refer to certain
 approximations and, for this reason,
can not be regarded completely consistent. It is evident that, in order to
be able to systematically study the nature of the scalar resonances, one should put the 
existing methods under the renewed scrutiny and look for rigorous
 criteria, which will not be based on the malevolent modifications of the
original theory.

In the present paper we make an attempt to answer the question, 
how the observables of the 
low-lying scalar mesons $a_0(980)$ and $f_0(980)$ can be (at least 
in principle) 
determined from lattice QCD simulations. It is clear that, due to the proximity
of the inelastic threshold, finite-volume effects should be very important and
could significantly distort the structure of the energy levels. 
Although the finite-volume corrections have been considered in
partially quenched ChPT at one loop (see, e.g.~\cite{Prelovsek:2010gm}), 
a systematic investigation of the problem, to the best of our knowledge,
 is still lacking. The present paper, in particular, intends to fill this gap.

In addition, we discuss, which conclusions (if any) 
about the nature of the scalar resonances $a_0(980)$ and $f_0(980)$
can be drawn {\em in a model-independent fashion}
from the calculations on the lattice. Namely, we reformulate the criterium
of Refs.~\cite{Weinberg:1963zz,Morgan:1992ge,Morgan:1993td,Tornqvist:1994ji}
for energy spectrum of lattice QCD in a finite box, whose volume-dependence
can be studied within the L\"uscher
framework~\cite{Luscher:1990ux}. In order to achieve the goal,
using the so-called twisted boundary condition~\cite{twisted} has
proven to be advantageous. We further investigate 
the relation of our approach to the HBC method of Ref.~\cite{Suganuma:2005ds}.

Further, we discuss a criterium, which can be used
 to distinguish between the 
$q\bar q$ mesons and the 
tetraquarks (but which does not distinguish between 
the tightly bound tetraquarks
and the $K\bar K$ molecules). The criterium is based on 
the study of the {\em strangeness
content} of these states. By using Feynman-Hellman theorem, this quantity
can be related to the quark mass dependence of the exotic state masses and
thus can be measured on the lattice.

The layout of the paper is as follows. In section~\ref{sec:Flatte} we briefly
review the phenomenological determination of the position of
 $a_0(980)$ and $f_0(980)$ poles in the complex plane and discuss the
``pole counting'' method. The generalization of this method to a finite
volume is considered in section~\ref{sec:finite}. In 
section~\ref{sec:s-content} we consider the strangeness content of the
scalar mesons and formulate a criterium for the
tetraquark candidates.
Section~\ref{sec:concl} contains our conclusions.

\section{Phenomenological analysis of the $\pi\pi$ scattering amplitude near $K\bar K$ threshold}
\label{sec:Flatte} 
In the following, we restrict ourselves to the discussion of the $f_0(980)$.
The case of the $a_0(980)$ can be considered along a similar path.

It is a well-known fact that $\pi\pi$ scattering below $K\bar K$
threshold is almost elastic (the inelasticity parameter is close to unity
in this region). For this reason, in the vicinity of the $K\bar K$
threshold, where the $f_0(980)$ resonance is located,
it is convenient to parameterize the $\pi\pi$ scattering amplitude in terms
of the coupled-channel $K$-matrix with the following 2-particle channels:
``1''=$\pi\pi$ and ``2''=$K\bar K$. 
The coupled-channel $T$-matrix in the S-wave
obeys the equation
\eq\label{eq:LS}
T_{ij}(s)=K_{ij}(s)+\sum_n K_{in}(s)\, iq_n(s)\, T_{nj}(s)\, ,\quad\quad
i,j,n=1,2\, ,
\en
where $q_1(s)=\sqrt{s/4-M_\pi^2+i0}$, $q_2(s)=\sqrt{s/4-M_K^2+i0}$
and $s$ denotes the
pertinent Mandelstam variable. Note that a similar
equation has been used in our treatment of the $\bar KN$ scattering 
length~\cite{Lage:2009zv}. However, 
in difference to Refs.~\cite{Lage:2009zv,Bernard:2008ax},
we do not refer to the non-relativistic effective theory in the derivation of
Eq.~(\ref{eq:LS}). Consequently, the functions $K_{ij}(s)$ are no more
 assumed to be
low-energy polynomials in the whole interval between $\pi\pi$ and $K\bar K$
thresholds. In principle, $K_{ij}(s)$ has a small imaginary part above
$4\pi$ threshold. However, in the following, we shall neglect this effect.

The S-wave $\pi\pi$ scattering amplitude is given by
\eq\label{eq:T11}
T_{11}(s)=\frac{1}{w_{11}^{-1}-iq_1}\, ,\quad\quad
w_{11}=K_{11}+\frac{iq_2\,K_{12}^2}{1-iq_2 K_{22}}\, .
\en
In the literature, different phenomenological parameterizations of the 
$K$-matrix are used. We distinguish between the parameterizations which have a
pre-existing real pole(s) in the vicinity of the $K\bar K$ threshold (see, e.g.,
Ref.~\cite{Au}) and those which are regular in this region 
(e.g.,~\cite{Protopopescu:1973sh,Oller:1997ng}). In the latter case, the 
$K$-matrix elements can be expanded in Taylor series 
$K_{ij}(s)=K^{(0)}_{ij}+q_2^2(s)\,K^{(1)}_{ij}+O(q_2^4)$,
whereas in the former, an additional pole term should be also included.
The location of the $S$-matrix pole(s) in either case is uniquely determined
by the behavior of the $K$-matrix in the threshold region. This defines
the two-step
strategy in the study of scalar mesons. In particular, we shall demonstrate
below that, measuring the energy spectrum in a finite volume, 
one may uniquely determine the $K$-matrix elements that amounts to measuring,
for instance, the coefficients $K^{(0)}_{ij},K^{(1)}_{ij}$ (this statement
is a generalization of L\"uscher method to the multichannel case). 
At the next stage, continuing the $K$-matrix into the complex plane by using
the Taylor (Laurent) expansion, one finds the location of the $S$-matrix poles
in the threshold region from the secular equation
\eq\label{eq:secular1}
1-iq_1(s)K_{11}(s)-iq_2(s)K_{22}(s)-q_1(s)q_2(s)(K_{11}(s)K_{22}(s)-K_{12}(s)^2)=0\, .
\en

Assuming further that the quantity $w_{11}^{-1}$ in Eq.~(\ref{eq:T11}) can be expanded in Taylor series in the variable $q_2$, one arrives at the so-called Flatt\'e
parameterization~\cite{Flatte:1976xu}
which modifies
the usual Breit-Wigner parameterization 
when a nearby threshold is present
\eq\label{eq:T11-exp}
T_{11}&=&\frac{\mbox{const}}{m_R^2-s+O(q_2^4)-im_R(g_\pi\,\hat q_1+g_K\,q_2(s)+O(q_2^2))}\, ,
\en
where $\hat q_1=q_1(s)\biggr|_{s=4M_K^2}$ and the parameters $m_R,g_\pi,g_K$ 
can be expressed through the quantities $K_{ij}^{(0)},K_{ij}^{(1)}$. However,
it turns out that, for the known phenomenological parameterizations, the Taylor expansion of
$w_{11}^{-1}$ has a very small radius of convergence. For this reason, 
it is safer to use the secular equation, Eq.\ref{eq:secular1}
to find the location of the poles.

The compositeness 
criterium~\cite{Weinberg:1963zz,Morgan:1992ge,Morgan:1993td,Tornqvist:1994ji} 
states that, for the molecular states,
only one of the poles is located in the vicinity of threshold, whereas
the nonmolecular state correspond to a pair of poles, both lying in the proximity
of the threshold. In terms of Flatt\'e parameters, the situation $g_K\gg g_\pi$
corresponds to the $K\bar K$ molecular state and vice versa.

To summarize, it is clear that the quantities to be measured in the
lattice simulations are the $K$-matrix elements in the vicinity of the
$K\bar K$ threshold.
 These, in turn, determine the position of the poles in the
$S$-matrix, that is, the energy and the width of $f_0(980)$.  
Consequently one can
 answer the question, whether $f_0(980)$ is a molecular state or not.
If the answer is negative this framework does not allow to further elaborate on 
the structure of this state. For that the method discussed in Section 4
has to be used.

\section{$K$-matrix formalism in a finite volume}
\label{sec:finite}

In Ref.~\cite{Luscher:1986pf} it is shown that the elastic 
scattering length can be extracted from lattice data, studying the volume
dependence of the ground state energy. This result is readily obtained by
Taylor expansion of 
a more general formula, which enables one to evaluate the elastic scattering
phase from the finite-volume energy spectrum~\cite{Luscher:1990ux}. At 
threshold, this scattering phase is determined by the effective-range expansion
parameters (scattering length, effective range, etc), which are thus also 
obtained from the analysis of the lattice data. Moreover, from the phase shift
determined on the lattice, one may (in principle) extract the position and
width of the elastic resonances. A further generalization of the approach
allows one to address the measurement of the resonance
formfactors~\cite{Hoja:2010fm}. For an alternative method to directly extract
the resonance position in the complex plane from the measured two-point function at finite times, see Ref.~\cite{Polejaeva}.

In the literature there have been attempts to generalize L\"uscher approach
to the case of the inelastic scattering (see, e.g.~\cite{Lage:2009zv,He}).
In particular, in Ref.~\cite{Lage:2009zv}, 
the use of a volume-dependent spectrum for the 
determination of the $K$-matrix elements, which are real quantities
was proposed.
The (complex) scattering length can be then expressed through these $K$-matrix
elements.

The main difference between the approach, Ref.~\cite{Lage:2009zv} and the
present one consists in the fact that in the former only the periodic
boundary conditions have been used. For this reason, it was not possible
to determine all three quantities $K_{ij},~i,j=1,2$ only from the data taken
exactly at one energy. Although the effects, coming from the momentum
dependence of $K_{ij}$ are power-suppressed at large volumes, they still
can represent a source of error, if the volume is not very large. In the
present paper we show that using twisted boundary conditions allows one
to circumvent this problem. 

Before formulating the approach in a finite volume, few remarks are in order.

\begin{itemize}

\item[i)]
The equation, which determines the finite-volume spectrum, has been obtained
in Refs.~\cite{Lage:2009zv,Bernard:2008ax} by using the non-relativistic effective theory.
This is the easiest and the most transparent way of derivation which, however,
implicitly assumes that the potential (or the $K$-matrix) is a low-energy
polynomial. In other words, it is assumed that the Taylor 
expansion of the potential
in the momentum space in powers of the relative 3-momenta converges
for all energies of interest.

The above requirement is certainly too restrictive, if one considers $\pi\pi$
scattering up to 1~GeV. On the other hand, this requirement is also 
superfluous. What is required, is that the effective potentials,
obtained by the 3-dimensional reduction of the coupled-channel 
$\pi\pi-K\bar K$ Bethe-Salpeter equation, are volume-independent up
to exponential corrections. In fact, this is the path of reasoning,
adopted in the original papers by 
L\"uscher~\cite{Luscher:1986pf,Luscher:1990ux}.

The generalization of the arguments of 
Refs.~\cite{Luscher:1986pf,Luscher:1990ux} to the coupled channel scattering
is relatively straightforward. To this end, first consider the (fictitious)
situation where $2M_\pi>M_K$. Then, 2-pion and kaon-antikaon states are
two states with the lowest energy. The 2-particle 
irreducible Bethe-Salpeter kernels $U_{ij}(P;p,q),~i,j=1,2$ 
in the infinite volume are analytic functions
of the center-of-mass energy 
$P_0$ in the range~$2M_\pi<P_0<4M_\pi$.
Then, the regular summation theorem~\cite{Luscher:1986pf}
implies that the finite-volume corrections to these kernels
at large volumes vanish faster than any inverse power of the volume.

Finally, one should perform analytic continuation in the pion mass. If both
pion and kaon masses are taken physical, inelastic thresholds move below
the $K\bar K$ threshold. Strictly speaking, it is {\em not} true any more that
corrections to the Bethe-Salpeter kernels are exponentially suppressed in the energy
interval we are considering. However,
as already mentioned, the coupling to the inelastic channels is extremely weak: 
the inelasticity parameter $\eta\geq 0.98$ below $K\bar K$ threshold.
This means that the coupling to the inelastic channels can significantly
influence the spectrum obtained in the absence of these channels, if the 
two-particle and multiparticle energy levels in a given volume accidentally come very close to
each other. Neglecting this possibility, we 
expect that inelastic channels effectively
decouple and the kernels can be considered to be almost volume-independent.
We shall use this assumption below.

\item[ii)]
On the cubic lattice the rotational symmetry is broken down and mixing of all
partial waves occur. For the problem in question this mixing, however, 
is strongly suppressed. In order to see this, let us recall that if the cubic
symmetry is not broken, the S-waves mix with the partial waves with 
the orbital momentum $l\geq 4$~\cite{Luscher:1990ux}. 
Below 1~GeV, the partial wave with $l=4$
is however very small (see, e.g.~\cite{Pelaez:2004vs}). Using Eq.~(6.15)
of Ref.~\cite{Luscher:1990ux} and the parameterization of the G-waves
from Ref.~\cite{Pelaez:2004vs}, we have estimated the correction term arising
from the mixing. In the region of interest it does not exceed a few percent. In the 
following, we neglect the mixing altogether.
 
\end{itemize}

\begin{figure}[t!]
\begin{center}
\includegraphics[width=8.cm]{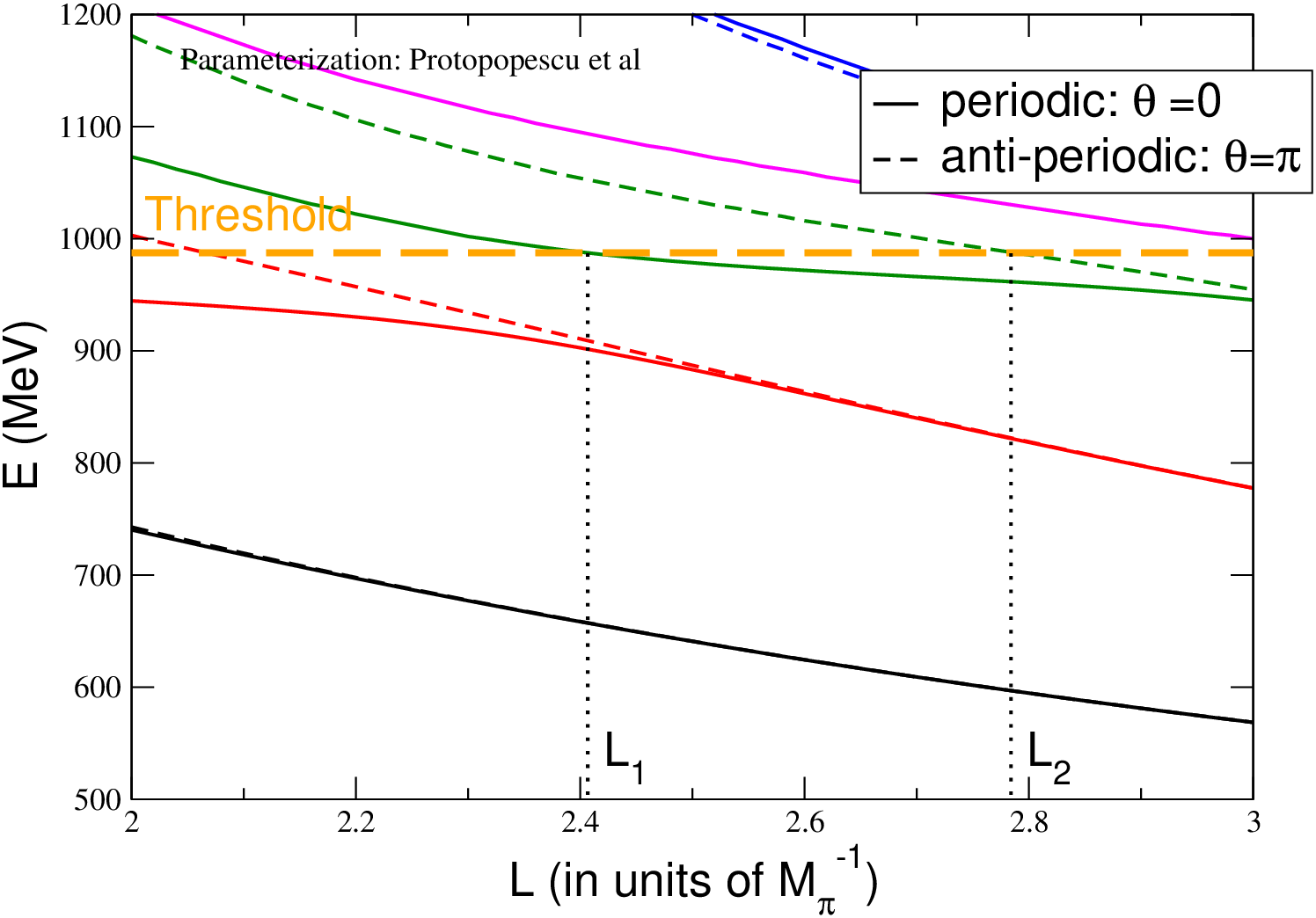}
\vspace*{1.cm}
\includegraphics[width=8.cm]{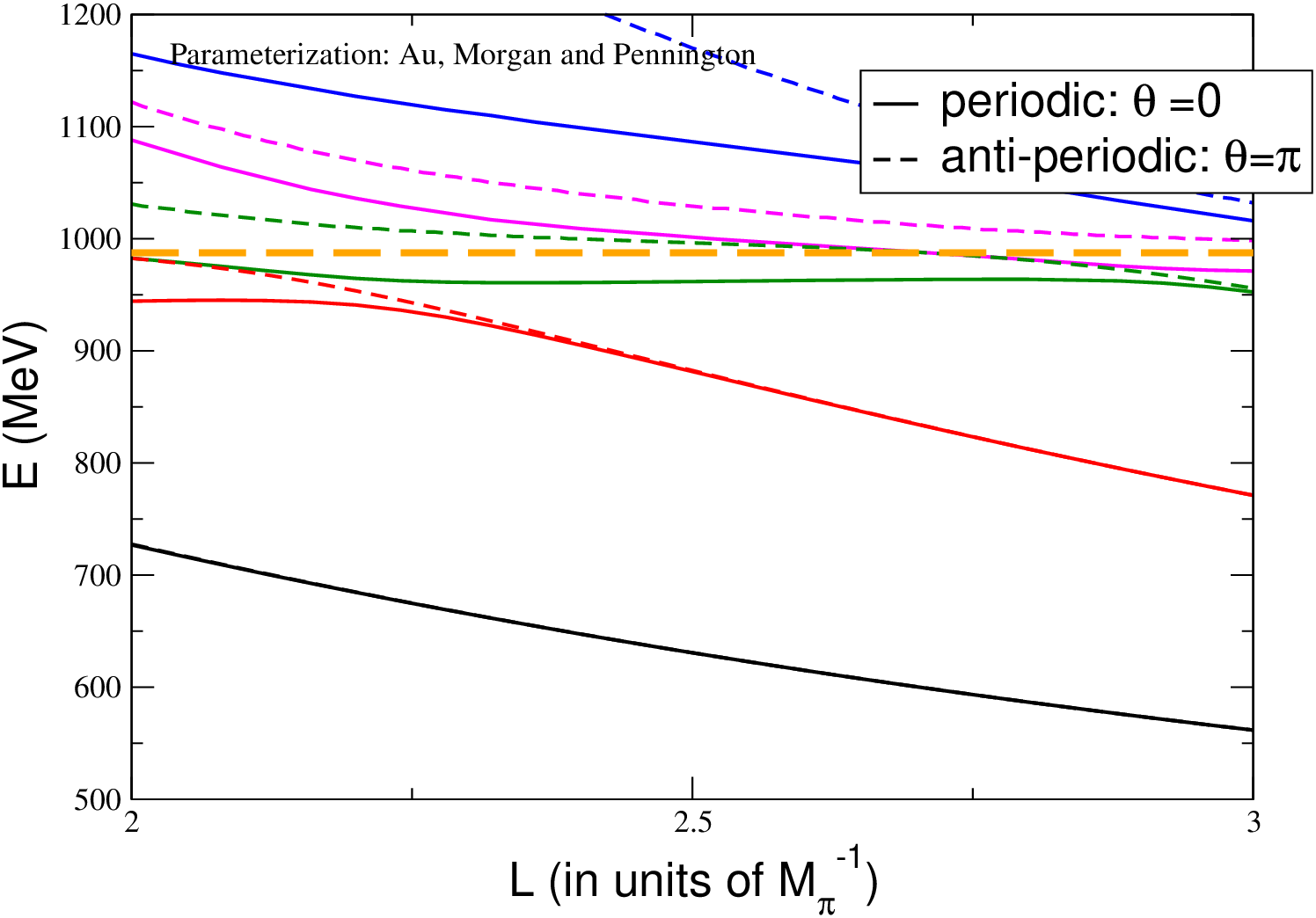}
\vspace*{1.cm}
\includegraphics[width=8.cm]{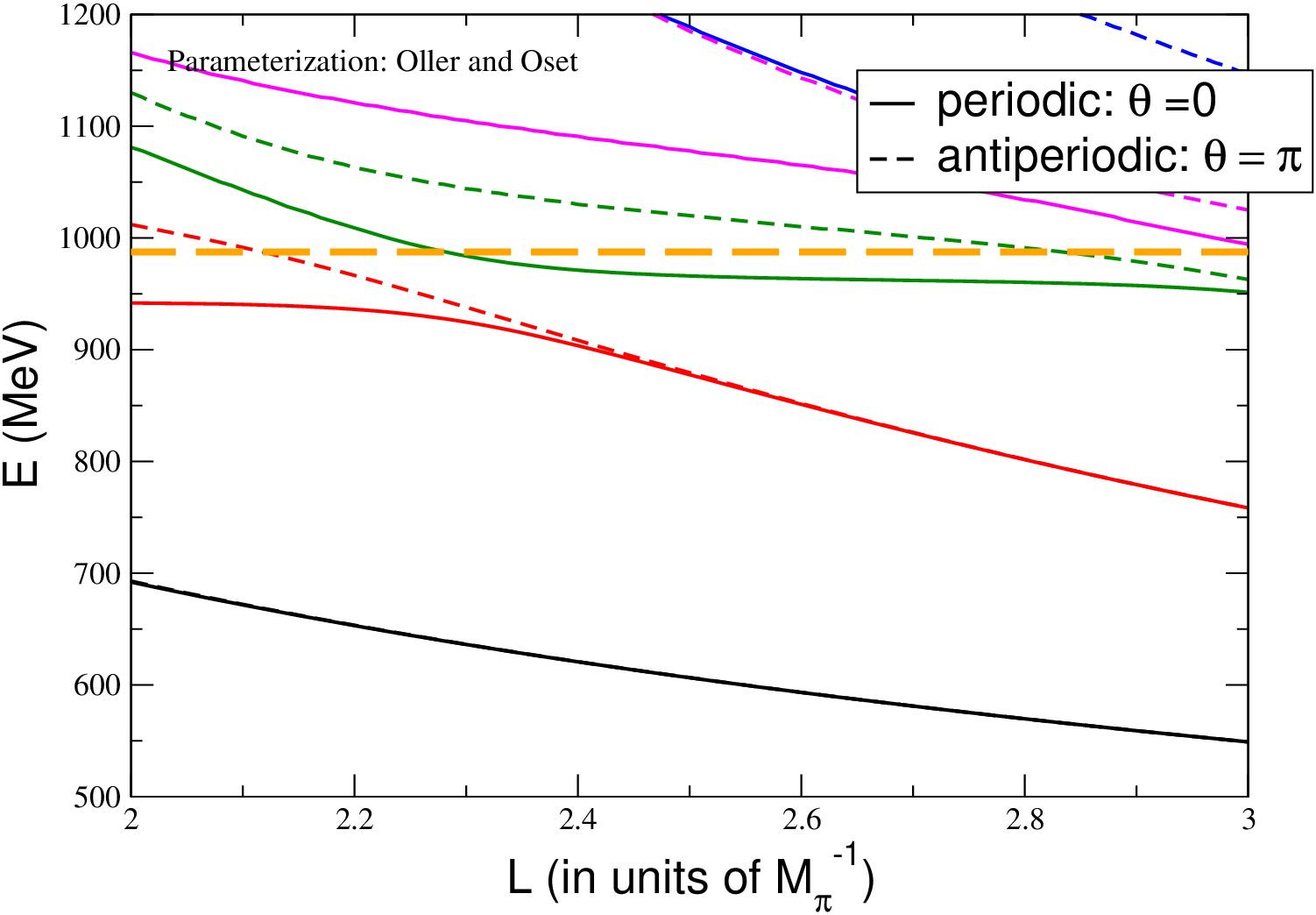}
\end{center}
\vspace*{-0.8cm}
\caption{Energy levels $E_n,~n=1,2,\cdots$ of the two-pion states in a 
finite box.
Solid lines correspond to the periodic boundary conditions ($\theta=0$),
and the dashed lines to the antiperiodic boundary conditions for the $s$-quark
($\theta=\pi$). Horizontal dashed line depicts the $K\bar K$ threshold.
Different parameterizations of the $K$-matrix from 
Refs.~\cite{Protopopescu:1973sh,Au,Oller:1997ng} have been used.
}
\label{fig:spectrum}
\end{figure}

\begin{figure}[t!]
\begin{center}
\includegraphics[width=8.cm]{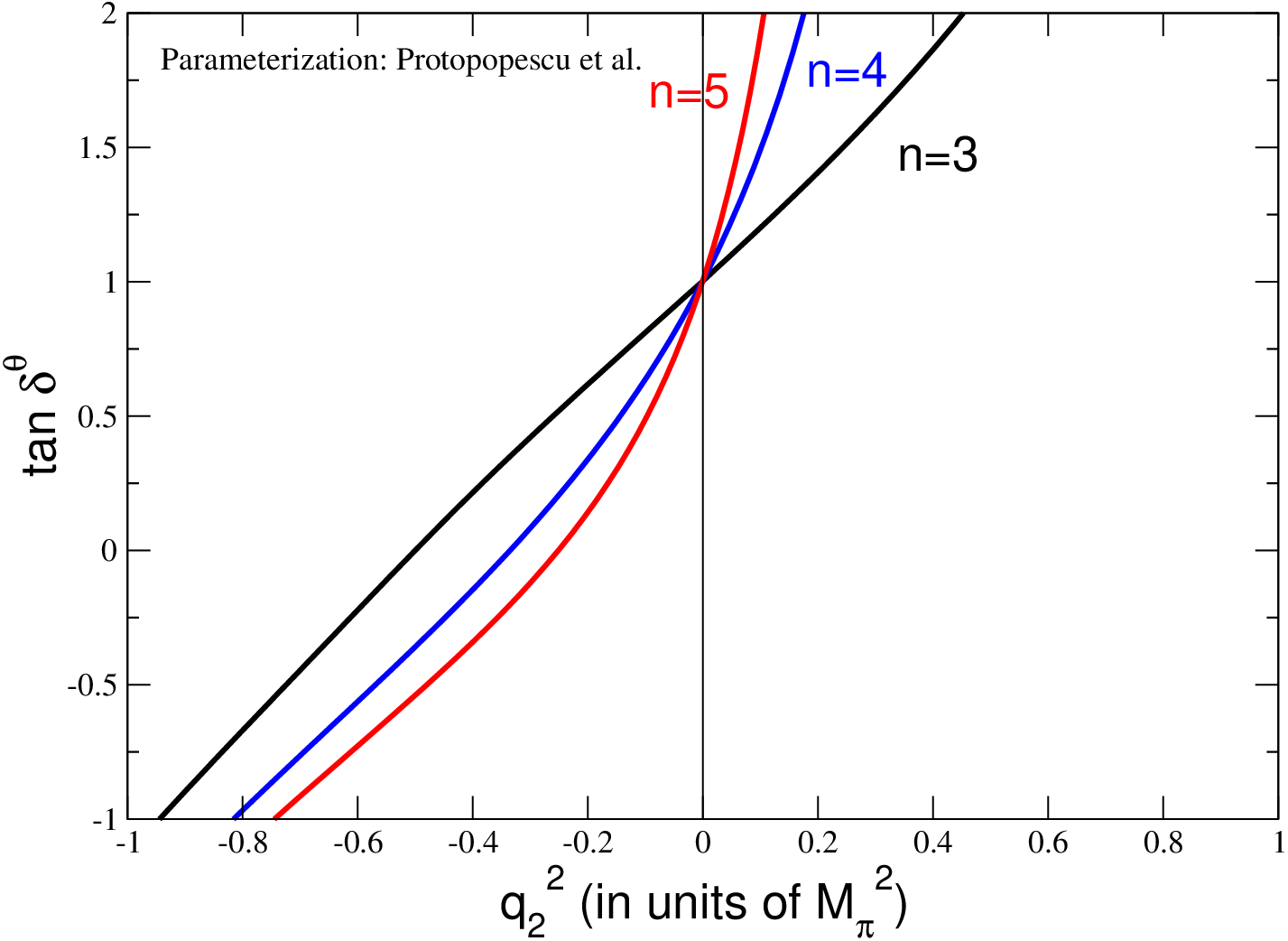}\hspace*{.3cm}
\vspace*{1.cm}
\includegraphics[width=8.cm]{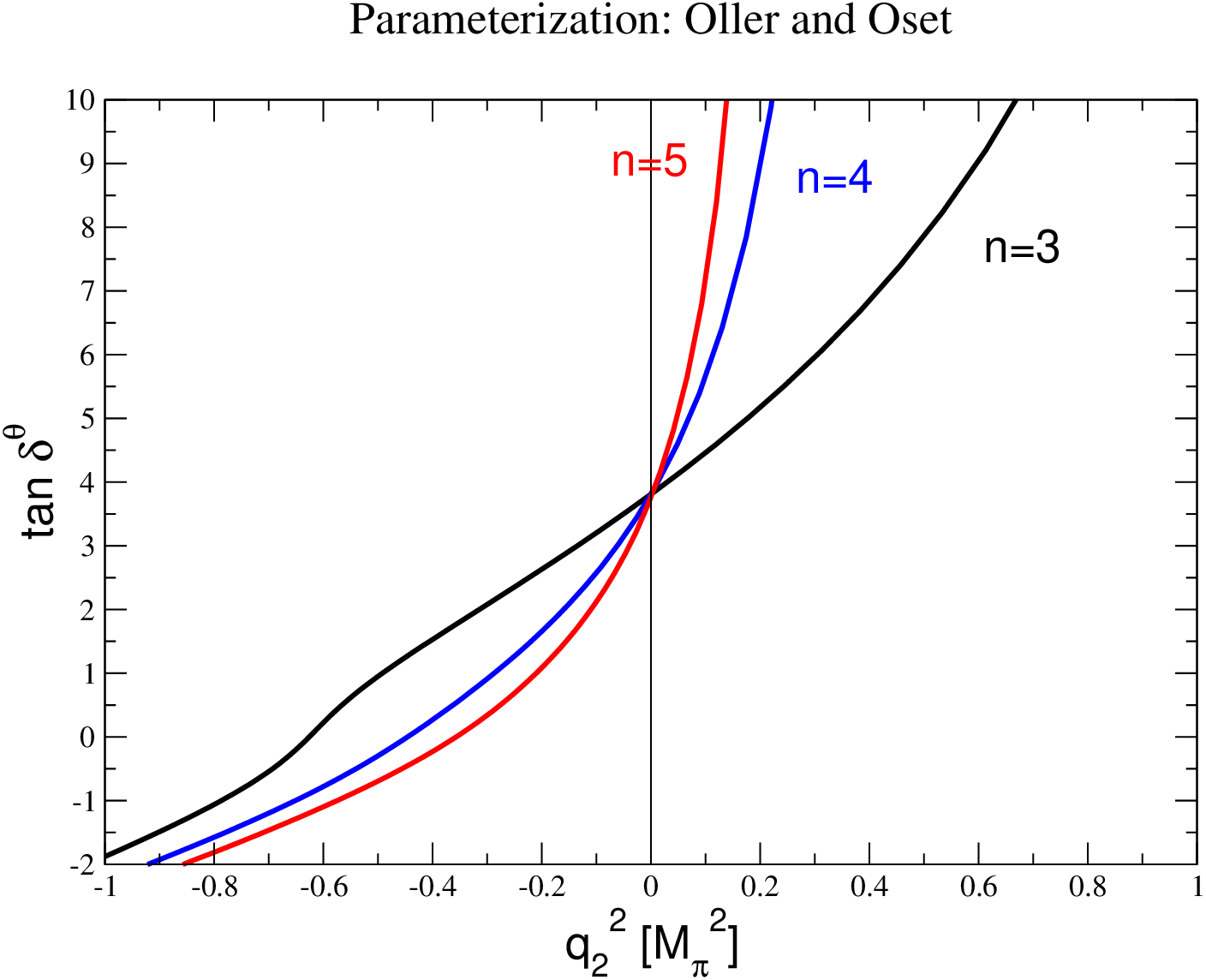}
\vspace*{1.cm}
\includegraphics[width=8.cm]{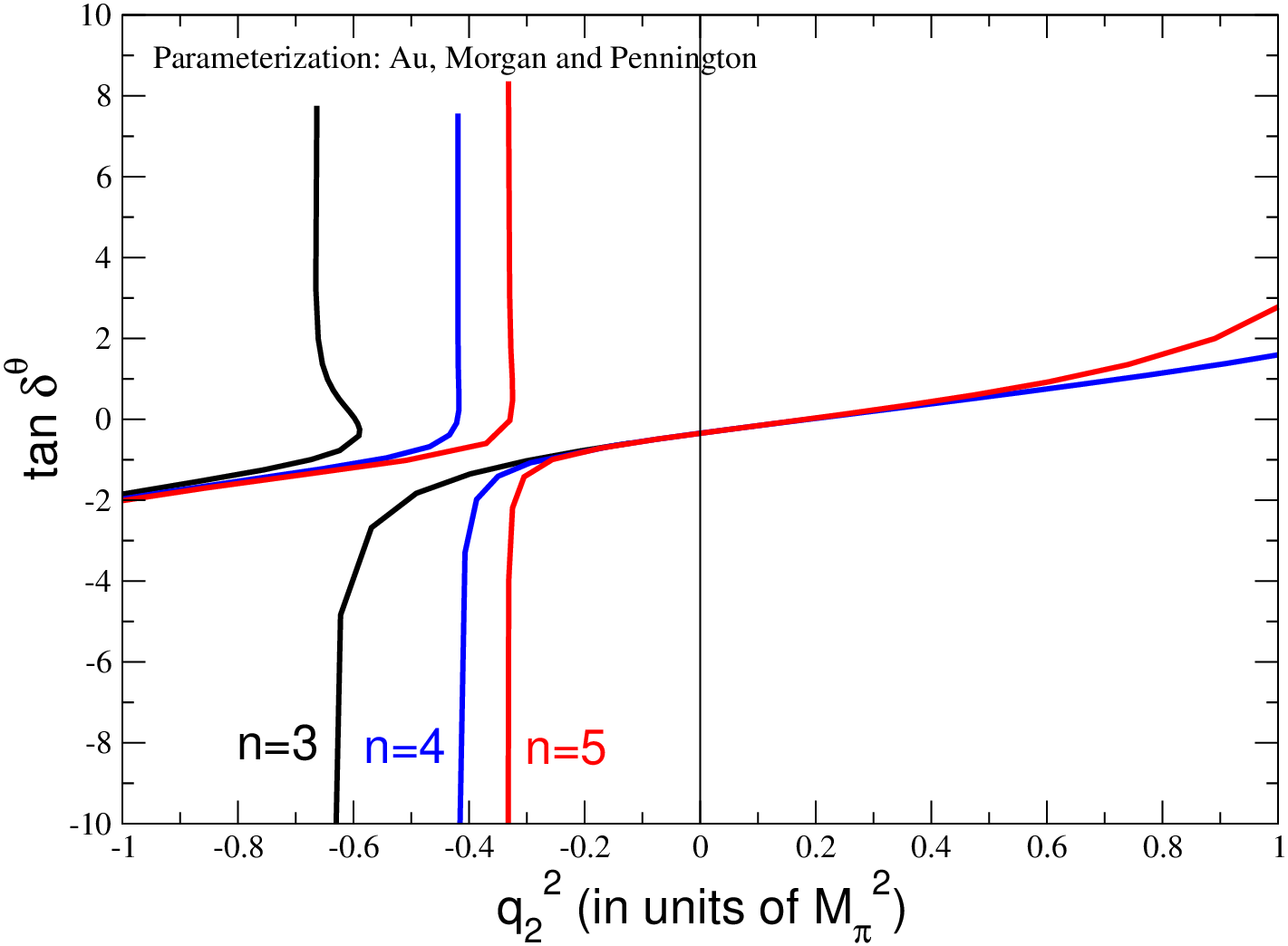}
\vspace*{-0.8cm}
\end{center}
\caption{The tangent of the pseudophase in the vicinity of
 the $K\bar K$ threshold, in case of the periodic boundary conditions
  ($\theta=0$). The pseudophase was extracted from the excited energy levels
$E_n$ with $n=3,4,5$. Different parameterizations have been used.}
\label{fig:pseudo-periodic}
\end{figure}

To summarize, with the above assumptions, the generalization of 
Eq.~(\ref{eq:LS}) to the finite-volume case is straightforward, and the result
is similar in form to that given in Ref.~\cite{Lage:2009zv}:
\eq\label{eq:LS_L}
T_{ij}(s)=K_{ij}(s)+\frac{2}{\sqrt{\pi} L}\,
\sum_n K_{in}(s)\, Z_{00}(1,k_n^2)\, T_{nj}(s)\, ,\quad\quad
i,j,n=1,2\, ,
\en
where $k_n(s)=\dfrac{q_n(s) L}{2\pi}$ and $Z_{00}(1;k^2)$ stands for L\"uscher 
zeta-function~\cite{Luscher:1986pf,Luscher:1990ux}.

The equation~(\ref{eq:LS_L}) implies periodic boundary conditions on both
pion and kaon fields. Below, we explore the possibility of using 
the so-called twisted boundary conditions, which we opt to impose only
on the strange quark field, whereas $u$- and $d$-quarks obey periodic
boundary conditions
\eq\label{eq:tbc}
u({\bf x}+L{\bf e}_i)=u({\bf x})\, ,\quad
d({\bf x}+L{\bf e}_i)=d({\bf x})\, ,\quad
s({\bf x}+L{\bf e}_i)=\mbox{e}^{i\theta}s({\bf x})\, ,\quad\quad
0\leq\theta<2\pi\, .
\en
In this relation, ${\bf e}_i,~i=1,2,3$ denote unit vectors along the
lattice axes, and $V=L^3$ is the spatial volume of the lattice. The vacuum
angle theta is chosen the same in all spatial directions, in order to avoid
the breaking of the cubic symmetry. This choice, however,
may be changed, if needed.

In the effective theory, the angle theta will appear in the boundary condition
for the kaon field and not for the pion field. 
\eq\label{eq:K}
K^\pm({\bf x}+L{\bf e}_i)=\mbox{e}^{\mp i\theta}K^\pm({\bf x})\, ,\quad
K^0({\bf x}+L{\bf e}_i)=\mbox{e}^{-i\theta}K^0({\bf x})\, ,\quad
\bar K^0({\bf x}+L{\bf e}_i)=\mbox{e}^{i\theta}\bar K^0({\bf x})\, .
\en
If $\theta\neq 0$, the $K\bar K$ pair at rest has the minimum relative momentum
$\pm\biggl(\dfrac{\theta}{L},\dfrac{\theta}{L},\dfrac{\theta}{L}\biggr)$ and the energy 
$2\sqrt{M_K^2+\dfrac{3\theta^2}{L^2}}$. In other words, the $K\bar K$
threshold can be moved by adjusting $\theta$. Exactly this property makes the
twisted boundary conditions particularly useful to study scalar mesons, which
are located very close to this threshold.

Twisted boundary conditions can be straightforwardly implemented in the
two-channel L\"uscher equation~(\ref{eq:LS_L}). To this end, it suffices to
replace $Z_{00}(1;k_2^2)\to Z_{00}^\theta(1;k_2^2)$, where
\eq\label{eq:Z00t}
Z_{00}^\theta(1,k_2^2)=\frac{1}{\sqrt{4\pi}}\lim_{s\to 1}
\sum_{{\bf n}\in Z^3}\frac{1}{\left(\sum_{i=1}^3(n_i+\frac{\theta}{2\pi})^2\right)-k_2^2}\, ,
\en
whereas  $Z_{00}(1;k_1^2)$ stays put. 

The energy levels are given by the secular equation, which is obtained from
Eq.~(\ref{eq:LS_L})
\eq\label{eq:secular}
&&1-\frac{2}{\sqrt{\pi}L}\,Z_{00}(1;k_1^2)\,K_{11}(s)
-\frac{2}{\sqrt{\pi}L}\,Z_{00}^\theta(1;k_2^2)\,K_{22}(s)
\nonumber\\[2mm]
 &+&\frac{2}{\sqrt{\pi}L}\,Z_{00}(1;k_1^2)\,
\frac{2}{\sqrt{\pi}L}\,Z_{00}^\theta(1;k_2^2)\,(K_{11}(s)K_{22}(s)-K_{12}(s)^2)
=0\, .
\en

Our aim is to describe a procedure, which enables one to extract the
$K$-matrix elements $K_{ij}$ in the vicinity of threshold from the lattice.
We shall illustrate this procedure on the example
of  the {\em synthetic data}, which were produced, using
different phenomenological parameterizations for the $\pi\pi$ S-wave phase shift and
the Eq.~(\ref{eq:secular}) to determine the spectrum for various values
of $\theta$.  We have tested the parameterizations, 
given in Refs.~\cite{Protopopescu:1973sh,Au,Oller:1997ng} (these are shortly described in
appendix~\ref{app:Protopopescu}), and produced the spectrum which is shown
in Fig.~\ref{fig:spectrum}. In this figure, the spectra at $\theta=0$ (solid
lines) and $\theta=\pi$ (dashed lines) are displayed. These spectra show quite
similar behavior despite the fact that the pertinent $K$-matrices have very
different properties. For example, the $K$-matrix from Ref.~\cite{Au} has a 
real pole very close to the $K\bar K$ threshold, whereas the $K$-matrix from Refs.~\cite{Protopopescu:1973sh,Oller:1997ng} is regular near threshold.
As expected, the spectrum is almost independent on the twist parameter $\theta$
away from threshold. Maximal variation is introduced in the vicinity of the threshold where the rearrangement of the levels occurs: the levels with 
$\theta=\pi$ for small values of $L$ 
are ``pushed up'' one level high as compared to the case with $\theta=0$.
Consequently, the measurements for different values of $\theta$
provide an independent piece of information in the threshold region, where the 
$\theta$-dependence is maximal. Since we are looking for 
the resonances 
exactly in this region, twisted boundary conditions can be used to
fix the location of these resonances. On the other hand, it is also clear 
from Fig.~\ref{fig:spectrum} that the attempts to identify separate levels with
either the resonance or scattering states are not very informative.
In particular, tuning the parameters $L,\theta$, one may easily move a single
energy level above or below the threshold.

In order to facilitate the extraction of the $K$-matrix elements from the data,
according to Ref.~\cite{Lage:2009zv}, it is convenient to define 
the pseudophase
\eq\label{eq:pseudophase}
\tan\delta^\theta(q_1)=-\tan\phi(k_1)\, ,\quad\quad
k_1=\frac{q_1L}{2\pi}\, ,\quad
\tan\phi(k_1)=-\frac{\pi^{3/2}k_1}{Z_{00}(1;k_1^2)}\, .
\en
The ($\theta$-dependent) energy spectrum $q_1=q_1(L)$ can be (in principle)
measured on the lattice. Consequently, the right-hand side of
Eq.~(\ref{eq:pseudophase}) (and, hence, the pseudophase) are measurables.
The physical meaning of the pseudophase is the following: apply L\"uscher 
formula
to the measured energy spectrum, assuming that the scattering is elastic
(in our case, this means that only $\pi\pi$ threshold is taken into account).
Thus, below the inelastic $K\bar K$ threshold, the pseudophase coincides with
the usual scattering phase. This is not the case above the inelastic threshold.

From Eq.~(\ref{eq:secular}) one
 may express the pseudophase through the $K$-matrix elements as follows:
\eq\label{eq:main}
 \tan\delta^\theta(q_1)=q_1\biggl(K_{11}(s)+
\frac{\frac{2}{\sqrt{\pi}L}\,Z_{00}^\theta(1;k_2^2)\,K_{12}(s)^2}
{1-\frac{2}{\sqrt{\pi}L}\,Z_{00}^\theta(1;k_2^2)\,K_{22}(s)}\biggr)\, .
\en
Suppose, for a moment, that $\theta=0$ and we tuned the box size so that
the energy is exactly equal to $2M_K$ (this corresponds to $L=L_1$ in
Fig.~\ref{fig:spectrum}) and measured the pseudophase for this box size.
Recalling that $Z_{00}(1;k_2^2)=-\dfrac{1}{\sqrt{4\pi}k_2^2}-2.514488997+O(k_2^2)$ as
$k_2^2\to 0$,  from Eq.~(\ref{eq:main}) we readily get 
\eq
\lim_{k_2^2\to 0}\tan\delta^\theta(q_1)\biggr|_{\theta=0}=\hat q_1
\biggl(K_{11}^{(0)}-\frac{(K_{12}^{(0)})^2}{K_{22}^{(0)}}\biggr)\, .
\en
This equation gives one relation between three quantities $K_{ij}^{(0)}$.

We repeat this procedure for a different value $L$ and adjust the
parameter $\theta$ so that energy of the measured level is 
exactly $2M_K$ again. For example in Fig.~\ref{fig:spectrum} this corresponds to $L=L_2$ at 
$\theta=\pi$. After performing three measurements at threshold energy and 
different values of $L,\theta$, we get enough equations to determine all  
$K_{ij}^{(0)}$ separately. Moreover, there is nothing special about
the threshold energy: the same procedure can be repeated at $\sqrt{s}\neq 2M_K$,
scanning the matrix elements $K_{ij}(s)$ in the vicinity of threshold.
In this way, one may e.g., answer the question, whether the $K$-matrix contains the
pre-existing poles in the threshold region.

\begin{figure}[t!]
\begin{center}
\includegraphics[width=8.cm]{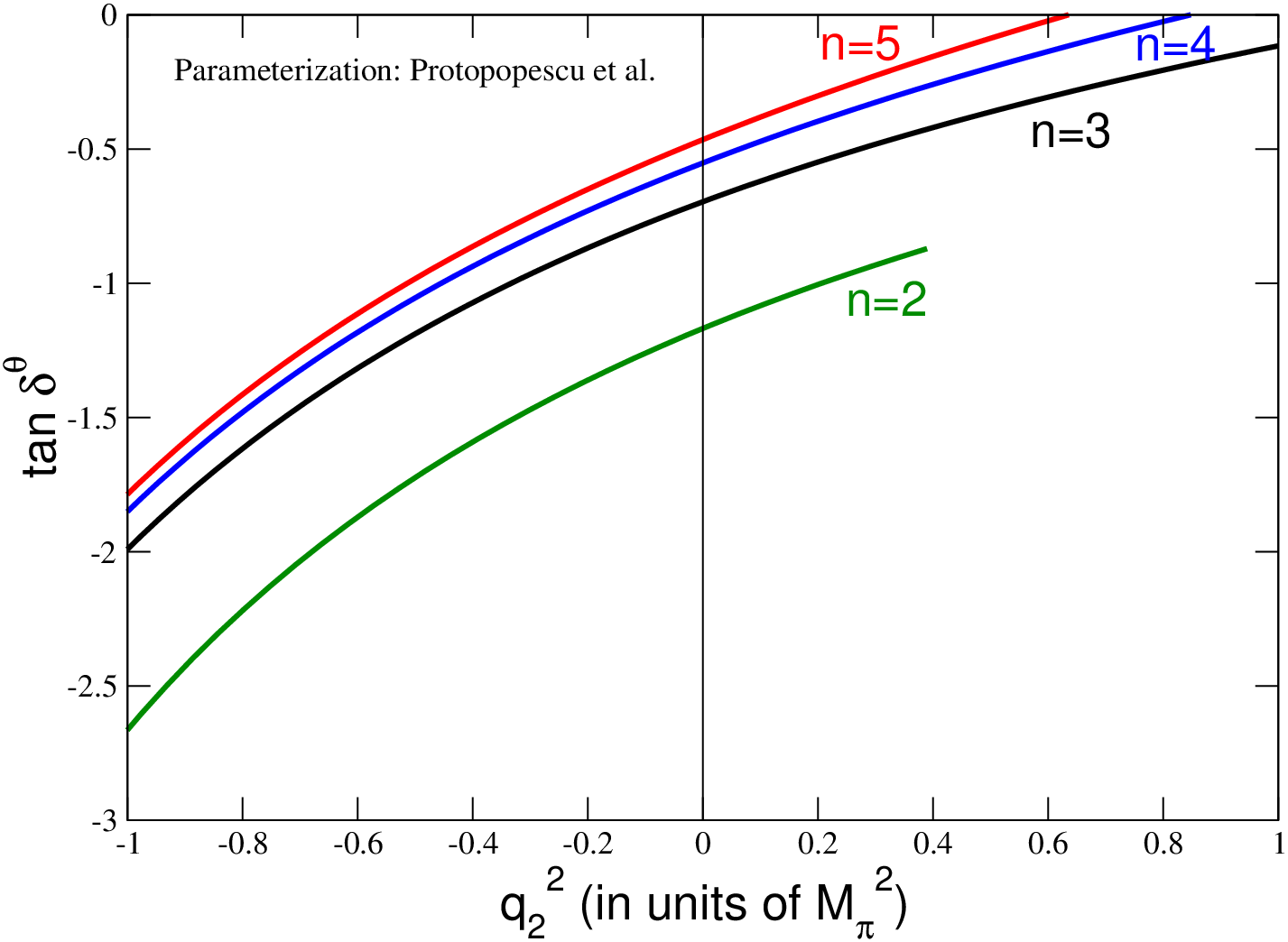}\hspace*{.3cm}
\vspace*{1.cm}
\includegraphics[width=8.cm]{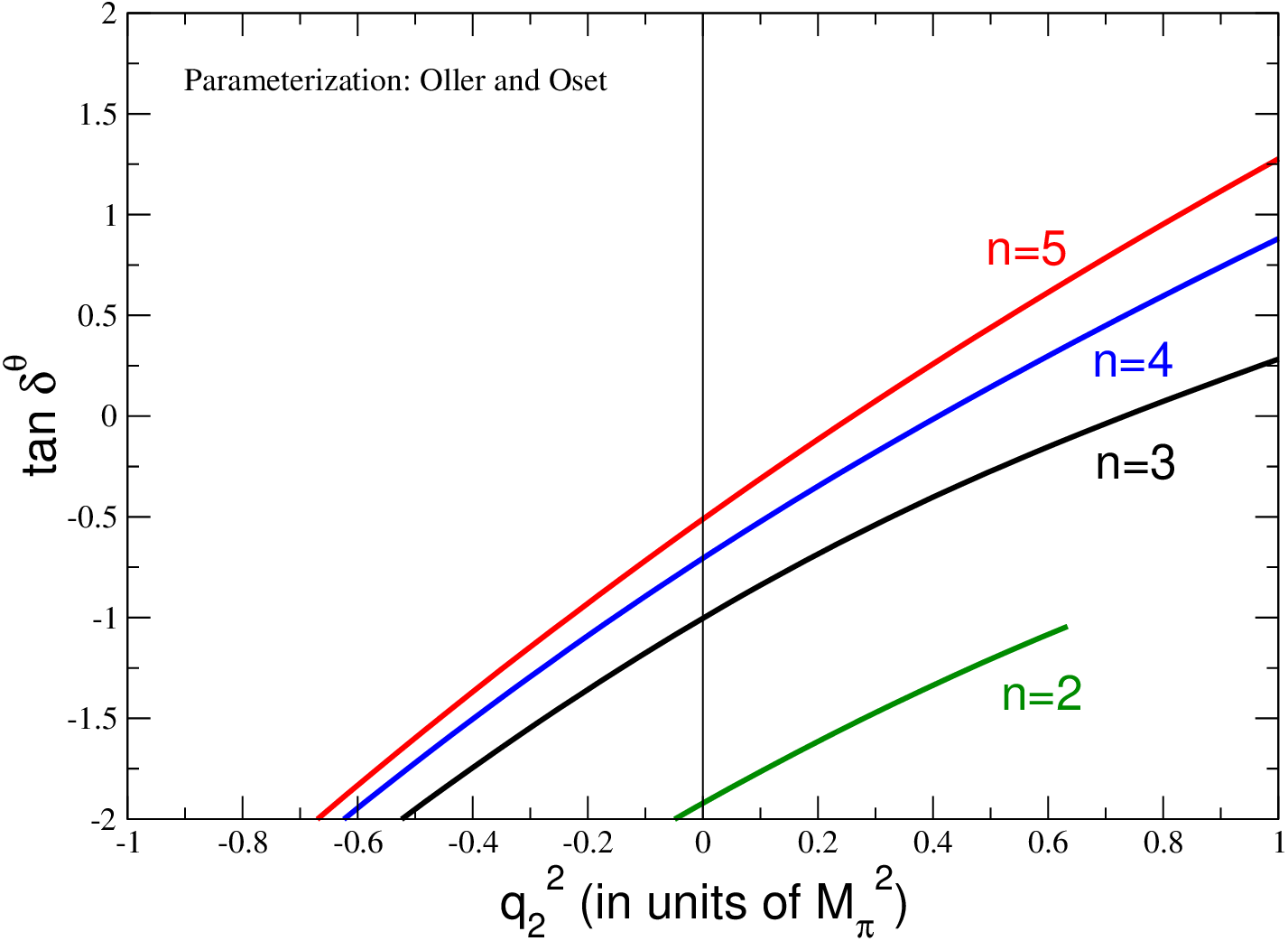}
\vspace*{1.cm}
\includegraphics[width=8.cm]{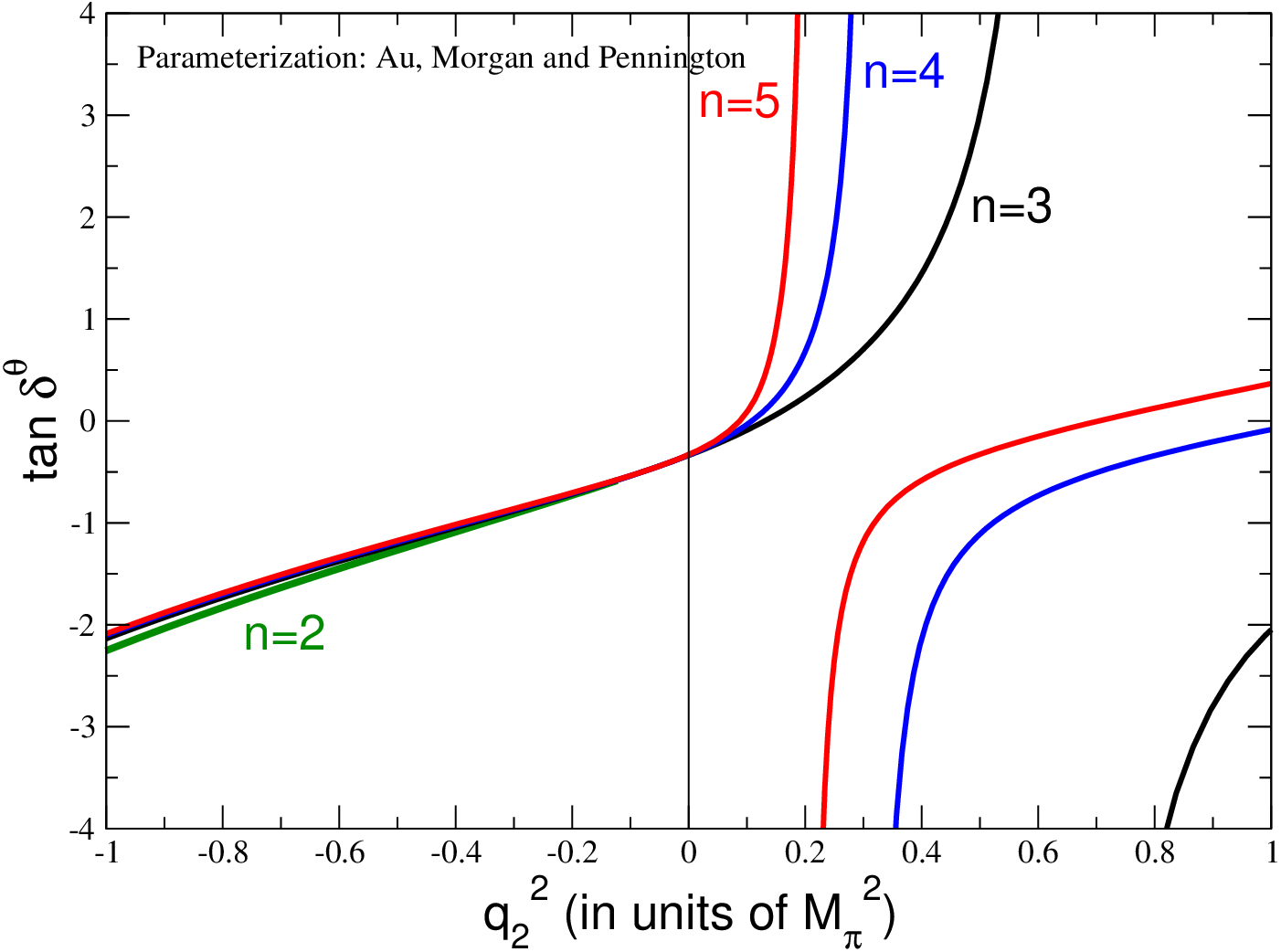}
\end{center}
\vspace*{-0.8cm}
\caption{The same as in Fig.~\ref{fig:pseudo-periodic}, but with antiperiodic
boundary conditions $\theta=\pi$.}
\label{fig:pseudo-antiperiodic}
\end{figure}

From Figs.~\ref{fig:pseudo-periodic} and \ref{fig:pseudo-antiperiodic}
one may conclude that the difference between different parameterizations
of the $K$-matrix is more clearly visible in the pseudophase that in the
structure of the energy levels which all show a similar behavior.
Moreover, it is seen that
the behavior of the pseudophase changes dramatically when $\theta$ 
changes from $0$ to $\pi$. On the basis of this observation one may
expect that the equations that relate the matrix elements $K_{ij}(s)$
with the measured pseudophases at the same energy are not degenerate and
will enable one to neatly extract these matrix elements from the data.

Last but not least, we wish to comment on the relation of the approach
 suggested
in the present paper with the method of HBC~\cite{Suganuma:2005ds}. 
We shall do this, adapting the argumentation of Refs.~\cite{Suganuma:2005ds}
to a choice of the boundary conditions used in the present article. 
Namely, as already mentioned above, 
varying the parameter $\theta$ from $0$ to $\pi$ amounts
to floating the $K\bar K$ threshold, whereas the $\pi\pi$ 
threshold stays put. Consequently, one interprets a given state as a 
$K\bar K$ scattering state if it is dragged along by the $K\bar K$ 
threshold, and as a ``genuine quark state,'' if it does not move.
On the other hand, from the expression of the pseudophase Eq.~(\ref{eq:main})
one may conclude that the energy level in the vicinity of the threshold does 
not move if the matrix element $K_{12}$ that describes the coupling of $\pi\pi$
and $K\bar K$ channels, is small. In terms of Flatt\'e parameters, this
corresponds to a small value of $g_K$, see Eq.\ref{eq:T11-exp}. 
In this case, a pair of poles appears in the
$S$-matrix near the $K\bar K$ threshold.
To summarize, it is seen that the approach proposed in the present paper
is a natural generalization of the  HBC method and enables one to extract
and to quantify the 
information about the nature of the resonances in the threshold region.

Finally, we wish to mention that we performed the calculation of the
energy levels and the pseudophases (for the different values of the
twisting angle $\theta$) for the case of the $a_0(980)$ resonance as well,
using the parameterization of Ref.~\cite{Oller:1997ng}. The results turn out  
to be qualitatively similar to the case of the $f_0(980)$.

\section{Strangeness content of the exotic states}
\label{sec:s-content}

As mentioned in the introduction, different methods are used at 
present to distinguish tet\-ra\-quarks from ordinary $q\bar q$ mesons in lattice 
QCD. Usually, a state is said to be a tetraquark if it is seen in the two-point
function of the operators $\bar qq\bar qq$ and not seen 
in the two-point function
of the operators $\bar q q$. More precisely, this means that the overlap of
the one particle state with the state produced from the vacuum by the 
operator $\bar qq\bar qq$ is much larger than with the state produced by 
the operator $q\bar q$. Furthermore, one may invoke arguments based on
the behavior of the spectrum in quenched approximation~\cite{Alford:2000mm}.
However, one should take these arguments with a grain of salt. 
For example, the the matrix elements used in the above argumentation
are scale-dependent and, in addition, depend on the way the
operators are constructed from the quark fields. It is clear that a 
mathematically consistent and model-independent criterium would be
highly desirable.

It should be pointed out that, generally speaking, the question whether
the composition of a particular state is $q\bar q$ or $q^2\bar q^2$, makes sense
e.g., in the non-relativistic quark models but not in the field theory. In the
latter, any operator with appropriate quantum numbers can be used as an 
interpolating field for a given particle. Thus, we have to look for a criterium
formulated in terms of {\em observable quantities,} which in the 
non-relativistic limit is coherent with our understanding of ordinary 
$q\bar q$ and tetraquark states. We shall use this strategy in the following.

Note first 
that the quark model wave functions of the states in the non-relativistic
limit are eigenfunctions of the operator
\eq\label{eq:op}
{\bf S}_i=\int d^3{\bf x}\,:\bar q_i(x)q_i(x):\, ,\quad\quad
i=u,d,s
\en
with the eigenvalues $N_i+N_{\bar i}$, where $N_i$ and $N_{\bar i}$ denote the
number of quarks and antiquarks of a flavor $i$ which are contained in this
state.

Consider now the $SU(3)$ nonets with maximal
mixing for the $q\bar q$ and $q^2\bar q^2$ mesons (the wave functions of the  
$q^2\bar q^2$ states are given, e.g., in Ref.~\cite{Polosa}).
Further, define the {\em strangeness content} of a state $B$ that belongs
to these nonets in a standard manner
\eq\label{eq:s-content}
y_B=\frac{2\langle B|\bar ss|B\rangle}{\langle B|\bar uu+\bar dd|B\rangle}\, .
\en
In the non-relativistic case, in order to calculate $y_B$, one has just to
count the number of the quarks (antiquarks) of a given species in a state
$B$. Using the wave functions from table~\ref{tab:s-content}, we may easily
evaluate the values of $y_B$. The results are given in the same table.

\renewcommand{\arraystretch}{1.8}
\begin{table}[t]
\begin{center}

\begin{tabular}{|l| c l| c l|}
\hline\hline
$B$ &\multicolumn{2}{|c|}{$q\bar q$} & \multicolumn{2}{|c|}{$q^2\bar q^2$}\\
\hline
$I=0$, nonstrange & $\frac{u\bar u+d\bar d}{\sqrt{2}}$, &$y_B=0$ &
$[ud][\bar u \bar d]$, & $y_B=0$ \\ 
$I=0$, strange & $s\bar s$, & $y_B=\infty$ & $\frac{[su][\bar s\bar u]+[sd][\bar s\bar d]}{\sqrt{2}}$, & $y_B=2$\\ 
$I=\frac{1}{2}$ & $u\bar s,~d\bar s\,+$ conj., & $y_B=2$ &
$[su][\bar u\bar d],~[sd][\bar u\bar d]\,+$ conj., 
& $y_B=\frac{2}{3}$\\ 
$I=1$ & $u\bar d,~\frac{u\bar u-d\bar d}{\sqrt{2}}\, ~d\bar u$, & $y_B=0$ &
$[su][\bar s\bar d],~\frac{[su][\bar s\bar u]-[sd][\bar s\bar d]}{\sqrt{2}} ,~
[sd][\bar s\bar u]$, & $y_B=2$\\
\hline\hline
\end{tabular}
\end{center}
\caption{The wave functions and the strangeness content of $q\bar q$ and $q^2\bar q^2$
mesons in the non-relativistic quark model with ideal mixing.}
\label{tab:s-content}
\end{table}

As one sees from table~\ref{tab:s-content}, the patterns followed by $y_B$ 
are very different for the $q\bar q$ mesons and tetraquarks. Note also that,
in case of arbitrary mixing, these two nonets can be still distinguished
due to the fact that $y_B$ for $I\neq 0$ does not depend on the mixing angle.

Next, we mention that the quantity $y_B$ is a well-defined quantity in QCD
(it is scale-independent) and can be directly evaluated on the lattice e.g.,
by studying the quark mass dependence of the scalar meson masses and applying
Feynman-Hellmann theorem
\eq\label{eq:FH}
y_B=2\biggl(\frac{dM_B}{dm_s}\biggr)\cdot\biggl(\frac{dM_B}{d\hat m}\biggr)^{-1}\, ,
\en
where we have assumed that isospin is conserved: $m_u=m_d=\hat m$.

Now we are in a position to formulate our proposal. The quantity $y_B$ is
a well-defined quantity in QCD and can be measured on the lattice.
On the other hand, in the non-relativistic quark models this quantity
clearly distinguishes between the $q\bar q$ and $q^2\bar q^2$ states. 
Therefore we
can use the measured pattern of the quantity $y_B$ for the nonet states
in order to {\em define}, what  we understand under tetraquark states
within relativistic quantum field theory. In contrast to the criteria used
in the literature so far, this definition, e.g.,
 does not operate with the quantities
that are scale-dependent (like the matrix elements of multiquark operators).   
Note that we took advantage here
of being able to vary quark masses on the lattice freely. 
Such a possibility is not available
in the phenomenological approaches based on the experimental input
and the above definition will be harder to use there.

Finally, note that the scalar mesons under consideration are resonances, not
stable particles. Since the width of these resonances is very small, 
it is natural to continue using Eqs.~(\ref{eq:s-content}) and 
(\ref{eq:FH}). For example, $M_B$ in Eq.~(\ref{eq:FH}) is to be now understood as 
the resonance pole position\footnote{The procedure of the extraction of 
the resonance matrix elements on the lattice is discussed, 
e.g., in Ref.~\cite{Hoja:2010fm}.}.
This means that the
measured energy levels should be first ``purified'' with respect to the finite
volume effects, as described in the previous sections. The method should be
applied at the final stage, to the resonance poles extracted from the spectrum.
If this is not done, the $K\bar K$ threshold, which will be moving if quark
masses are varied, could strongly influence nearby energy 
levels, and this may result in wrong
conclusions about the quark mass dependence of the true resonance energies. 

Interestingly, the kaon mass dependence of the scalar mesons can also be
used as a signature of the molecular picture. 
For a such a scenario, the leading
Fock component of the scalar meson wave function has two quarks and two
anti-quarks, much like the just discussed tetraquark states. However, while
these are expected to be compact, a molecule is loosely bound and thus
spatially extended. The molecular nature leads to a very peculiar kaon mass
dependence of the molecule as shown in Ref.~\cite{Cleven:2010aw}. The mass
of a $\bar KK$-molecule can be written as $M_{\rm mol}= M_K + M_{\bar K} -
\epsilon$, with $\epsilon$ the small binding energy, $\epsilon \ll M_K$.
Consequently, the kaon mass dependence of  $M_{\rm mol}$ is expected to be
{\it linear with a slope of two}, with correction of the 
order ${\cal O}(r^2 \epsilon M_{\rm mol)}$,
where $r$ denotes the range of forces. This is dramatically different from a
tetraquark, where the kaon mass dependence is generated either from the
strange valence or the  strange sea quarks and is thus expected to depend on $m_s$
linearly or as $m_s^{3/2}$, generating a leading kaon mass dependence 
as $M_K^2$ or $M_K^3$, respectively. For a more detailed discussion on this
issue, we refer to  Ref.~\cite{Cleven:2010aw}.

\newpage

\section{Conclusions}
\label{sec:concl}

The main results of this investigation can be summarized as follows:

\begin{itemize}

\item[i)]
Obviously, the measured excited spectrum can not be directly identified with
the experimentally observed scalar mesons. This simple fact becomes crystal
clear by looking at Fig.~\ref{fig:spectrum}, where one has freely used the parameters
$L$ and $\theta$ to move an energy level below or above $2M_K$. What defines
the energy and the width of a resonance is the position of the $S$-matrix pole
in the complex plane. This position can be determined, extracting $K$-matrix
elements from the spectrum measured on the lattice.

\item[ii)]
The procedure of determining $K$-matrix elements at the inelastic threshold,
which was described in the present paper, is the generalization of L\"uscher's
method to the elastic scattering length. It is also an improvement of the
method described in Ref.~\cite{Lage:2009zv}. Namely, using twisted boundary
conditions, as proposed in the present paper, enables one to determine all
three quantities $K_{ij}(s)$ in the vicinity of the inelastic threshold.

\item[iii)]
L\"uscher's approach implies the study of the response of the energy spectrum on
the variation of the box size $L$. 
We have shown that, for certain quantities,
studying the dependence on the twisting angle $\theta$ may partly 
substitute studying
the volume dependence.
In other words, one may use the twisting parameter $\theta$ to scan 
the energy region in the vicinity of the $K\bar K$ threshold.

\item[iv)]
The study of the $L$- and $\theta$-dependence of the energy spectrum of scalar mesons as
proposed in the present paper is beyond any doubt a very demanding
enterprise. The authors bear no illusion that the whole program can be
realized in lattice calculations at physical quark masses anytime soon,
especially for the $f_0(980)$ meson (the situation with $a_0(980)$ could be
slightly better). However, we still find it important to formulate a rigorous
way to treat the problem in question, which can be used one day.

\item[v)]
We show that the use of the twisted boundary conditions allows one to
distinguish between the loosely bound molecular states and the compact 
quark compounds. 
We in addition argue that if the latter possibility 
is realized, the measured values of the strangeness content 
for the different members of the $SU(3)$ nonet
allow one to interpret these states either as  conventional 
$\bar q q$ states or  $q^2\bar q^2$ tetraquark states. 
Thus the measurement of the strangeness content on the lattice, 
which can be achieved by
studying the quark mass dependence of the resonance energies,
enables one to gain detailed
information about the structure of the scalar mesons.
We have also pointed out that the molecular picture can be further tested 
from the measurement
of the kaon mass dependence of the mass of the scalar mesons - 
for a molecular state this would be linear with slope two 
(modulo small corrections).

\end{itemize}

\bigskip

{\em Acknowledgments.} We would like to thank V.~Baru, J.~Gasser, C.~Hanhart,
B.~Metsch, B.~Moussallam, J.~Oller, J.~Pelaez, S.~Prelovsek and C.~Urbach for useful discussions. 
This work is supported in part by DFG (SFB/TR 16,
``Subnuclear Structure of Matter'') and  by the Helmholtz Association
through funds provided to the virtual institute ``Spin and strong
QCD'' (VH-VI-231). We also acknowledge the support of the European
Community-Research Infrastructure Integrating Activity ``Study of
Strongly Interacting Matter'' (acronym HadronPhysics2, Grant
Agreement n. 227431) under the Seventh Framework Programme of EU.
A.R. acknowledges  support
of the Georgia National Science Foundation (Grant \#GNSF/ST08/4-401).


\renewcommand{\thefigure}{\thesection.\arabic{figure}}
\renewcommand{\thetable}{\thesection.\arabic{table}}
\renewcommand{\theequation}{\thesection.\arabic{equation}}

\appendix

\setcounter{equation}{0}
\setcounter{figure}{0}
\setcounter{table}{0}

\section{Different parameterizations of the two-channel\\ $K$-matrix}
\label{app:Protopopescu}

A particular parameterization of the $K$-matrix elements from the paper
Protopopescu {\it et al.,} Ref.~\cite{Protopopescu:1973sh}, 
which is used in the present paper, is given by 
\eq
K_{ij}=(M^{-1})_{ij}\, ,\quad\quad
M_{ij}(s)=M_{ij}^{(0)}+(s-s_t)M_{ij}^{(1)}\, ,\quad\quad s_t=4M_K^2\, ,
\en
where the coefficients $M_{ij}^{(n)}$ take the following values
\begin{center}
\begin{tabular}{|c|c|c|c|}
\hline
& $M_{11}^{(n)}$&$M_{12}^{(n)}$&$M_{22}^{(n)}$\\
\hline
$n=0$ & $3.38\,M_\pi$ & $2.40\, M_\pi$ & $0.071\, M_\pi$ \\
\hline
$n=1$ & 0 & 0 & $-0.0038\, M_\pi^{-1}$ \\
\hline
\end{tabular}
\end{center}
The scattering matrix in the paper by Oller and Oset, Ref.~\cite{Oller:1997ng},
is given by a solution of the 2-channel Bethe-Salpeter equation
\eq
T_{ij}(s)=V_{ij}(s)+\sum_n V_{in}(s)G_n(s)T_{nj}(s)\, ,\quad\quad
i,j,n=1,2\, ,
\en
where $V_{ij}(s)$ are the tree-level meson-meson scattering amplitudes, 
calculated in Chiral Perturbation Theory
\eq
V_{11}=-{\cal N}\,\frac{2s-M_\pi^2}{2F_\pi^2}\, ,\quad
V_{12}=V_{21}=-{\cal N}\,\frac{\sqrt{3}s}{4F_\pi^2}\, ,\quad
V_{22}=-{\cal N}\,\frac{3s}{4F_\pi^2}, \quad
{\cal N}=-(8\pi\sqrt{s})^{-1}
\en
with $F_\pi\simeq 93~\mbox{MeV}$
and the loop functions $G_k(s)$ are given by
\eq
G_n={\cal N}^{-1}\biggl\{-\frac{1}{8\pi^2}\,\ln\frac{q^{\sf max}}{M_n}
\biggl(1+\frac{w^{\sf max}_n}{q^{\sf max}}\biggr)
+\frac{\sigma_n}{16\pi^2}\,\ln\frac{\sigma_n w^{\sf max}_n/q^{\sf max}+1}
{\sigma_n w^{\sf max}_n/q^{\sf max}-1}\biggr\}\doteq G_n^{\sf R}+iq_n(s)\, .
\en
In the above expression, $\sigma_n=\sqrt{1-\frac{4M_n^2}{s}+i0}$,
$q^{\sf max}=\sqrt{\Lambda^2-M_K^2}$ 
is the cutoff momentum in the loops and  
$w^{\sf max}_n=\sqrt{M_n^2+(q^{\sf max})^2}$. The cutoff parameter $\Lambda$
is chosen to be $\Lambda\simeq 1020~\mbox{MeV}$. 

The $K$-matrix elements are given by a solution of the equation
\eq
K_{ij}(s)=V_{ij}(s)+\sum_n V_{in}(s)G_n^{\sf R}(s)K_{nj}(s)\, .
\en
In analogy to the parameterization by Protopopescu {\it et al.,} the above
$K$-matrix elements are regular in the vicinity of the $K\bar K$ threshold. The
resonance emerges due to the rescattering effect.

In difference to this, the parameterization of the $K$-matrix in the paper by
Au {\it et al.,}, Ref.~\cite{Au} contains the pre-existing pole in the vicinity
of the $K\bar K$ threshold. This parameterization looks as follows
\eq
K_{ij}=\frac{s-s_0}{4M_K^2}\,\biggl\{
\frac{f_if_j}{(s_1-s)(s_1-s_0)}+\sum_{n=0}^4c^n_{ij}
\biggl(\frac{s}{4M_K^2}-1\biggr)\biggr\}\, 
\en
where $s_0=-0.0162$, $s_1=0.9383$, 
$f_1=-0.1659$,
$f_2=0.5852$ and the values of the coefficients $c^n_{ij}$ are given by
\begin{center}
\begin{tabular}{|c|c|c|c|}
\hline
& $c_{11}^n$&$c_{12}^n$&$c_{22}^n$\\
\hline
$n=0$ &0.4247  &-3.1401 & -2.8447 \\
\hline
$n=1$ &-0.5822 &-0.1359 & 6.9164 \\
\hline
$n=2$ &2.5478 &1.0286 & 5.2846 \\
\hline
$n=3$ &-1.7387 &-2.3029 & -0.9646 \\
\hline
$n=4$ &0.8308 & 0.1944&  0\\
\hline
\end{tabular}
\end{center}
All dimensionful quantities are given in 
powers of GeV.

\end{document}